\documentclass[aps,prl,twocolumn,showpacs,floatfix,preprintnumbers,nofootinbib,superscriptaddress]{revtex4}

\usepackage{graphicx}
\usepackage{adjustbox}
\usepackage{color}
\usepackage{natbib}
\usepackage{multirow}
\usepackage{amsmath}
\usepackage[amssymb]{SIunits}
\usepackage{epstopdf}

\begin{document}

\title{Linear scale bounds on dark matter--dark radiation interactions
and connection with the small scale crisis of cold dark matter}

\author{Maria Archidiacono}
\affiliation{Institute for Theoretical Particle Physics and Cosmology (TTK), \\ RWTH Aachen University, D-52056 Aachen, Germany}
\author{Sebastian Bohr}
\affiliation{Institute for Theoretical Particle Physics and Cosmology (TTK), \\ RWTH Aachen University, D-52056 Aachen, Germany}
\author{Steen Hannestad}
\affiliation{Department of Physics and Astronomy, Aarhus University, 8000 Aarhus C, Denmark}
\author{Jonas Helboe J{\o}rgensen}
\affiliation{Department of Physics and Astronomy, Aarhus University, 8000 Aarhus C, Denmark}
\author{Julien Lesgourgues}
\affiliation{Institute for Theoretical Particle Physics and Cosmology (TTK), \\ RWTH Aachen University, D-52056 Aachen, Germany}

\date{\today}

\preprint{TTK-17-21}

\begin{abstract}
One of the open questions in modern cosmology is the small scale crisis of the cold dark matter paradigm.
Increasing attention has recently been devoted to self-interacting dark matter models as a possible answer. However, solving the so-called ``missing satellites" problem requires in addition the presence of an extra relativistic particle (dubbed dark radiation) scattering with dark matter in the early universe. Here we investigate the impact of different theoretical models devising dark matter dark radiation interactions on large scale cosmological observables. We use cosmic microwave background data to put constraints on the dark radiation component and its coupling to dark matter. We find that the values of the coupling allowed by the data imply a cut-off scale of the halo mass function consistent with the one required to match the observations of satellites in the Milky Way.

\end{abstract}

\maketitle

{\em Introduction}---
The existence of a non-relativistic non-baryonic component of the Universe, namely dark matter,
is one of the building blocks of modern cosmology and astrophysics. Large scale observations have set cold dark matter as the dominant paradigm for the interpretation of cosmological and astrophysical data.
The cold dark matter paradigm implies that the new non-relativistic particle is collisionless and interacts only through gravity. 

However simulations of pure cold dark matter are at odds with observations at small scales:
the satellites predicted by simulations are too many (``missing satellites" problem),
with a cuspy profile (``cusp vs core" problem) and with a larger circular velocity, i.e.,
larger enclosed mass, (``too big to fail" problem) compared to the dwarf galaxies gravitationally bounded to the Milky Way. 

Baryonic physics can mitigate some of these problems~\cite{Sawala:2015cdf},
however hydrodynamical simulations show that baryons cannot fully solve all the problems at once~\cite{Pawlowski:2015qta}.
 
Self-interacting dark matter (SIDM) models~\cite{Bringmann:2016ilk,Ahlgren:2013wba,Aarssen:2012fx,Cyr-Racine:2015ihg,Buckley:2009in,Buckley:2014hja,Loeb:2010gj,
Bellazzini:2013foa,Boddy:2014yra,Kainulainen:2015sva,Tulin:2013teo},
supplementary to baryonic physics, provide a solution to the last two problems.
Dark matter self-interactions arise from a secluded force, mediated by a new light boson.
The Yukawa potential related to the hidden force induces a velocity dependent cross section:
when the velocity dispersion is around the typical value of the relative velocity of dark matter particles in dwarf galaxies
($v_\mathrm{rel} \sim 10 \, \mathrm{km/s}$), the cross section shows a Sommerfeld-like enhancement.
The scatterings flatten the inner density profile and dissipate energy outwards, making the internal structure of the simulated satellites consistent
with the morphology of
the dwarf galaxies observed in the Milky Way (see Ref.~\cite{Vogelsberger:2015gpr} for the first SIDM simulation).

The solution of the ``missing satellites" problem requires the existence of a non-standard relativistic partner of dark matter,
namely dark radiation~\cite{Chu:2014lja,Diamanti:2012tg,Schewtschenko:2015rno,Buen-Abad:2015ova,Ackerman:mha}.
The scattering between the relativistic dark radiation particle and the non-relativistic dark matter particle keeps
the latter in thermal equilibrium until the kinetic decoupling, when the momentum transfer rate $\Gamma$ falls below
the expansion rate of the Universe
$H$~\footnote{Since $H$ is proportional to $T^2$ during radiation domination, 
dark radiation and dark matter decouple if $\Gamma \propto T^n$ with $n \geq 2$, otherwise they would {\it re-couple}.}.
Perturbations on scales smaller than the scale entering the horizon at the time of kinetic decoupling cannot grow;
therefore the matter power spectrum is suppressed on those small scales and the number of satellites is reduced.

The hypothesis of sterile neutrinos playing the role of the relativistic dark matter partner~\cite{Binder:2016pnr,Archidiacono:2014nda,
Wilkinson:2014ksa,Escudero:2015yka,Bringmann:2013vra,Dasgupta:2013zpn}
is of particular interest. Indeed, if sterile neutrinos are charged under the same new force, then the new dark sector can provide a
comprehensive solution, not only to the small scale crisis of cold dark matter, but also of the tension between neutrino oscillation anomalies
and cosmic microwave background (CMB) data~\cite{Archidiacono:2016kkh}.
Moreover, secluded sterile neutrinos~\cite{Chu:2015ipa,Hannestad:2013ana,Mirizzi:2014ama,Saviano:2014esa,Archidiacono:2015oma},
mixing with the active neutrinos, can potentially be constrained at IceCube~\cite{Cherry:2014xra,Arguelles:2017atb,Cherry:2016jol}.

This work is aimed at verifying
if cosmology is sensitive to the microphysics behind the phenomenological model
and if the CMB constraints on the coupling constant are consistent with the
astrophysical requirements to solve the small scale crisis.

{\it Methodology}---
We have modified the Boltzmann solver {\sc class}\footnote{\tt http://class-code.net}~\cite{Lesgourgues:2011re,Blas:2011rf,Lesgourgues:2011rh}
to include the very flexible ETHOS~\cite{Cyr-Racine:2015ihg} parametrization of dark radiation--dark matter interactions. Compared to the implementation of  ETHOS in {\sc CAMB}~\cite{Lewis:1999bs} that has been released by the authors of Ref.~\cite{Cyr-Racine:2015ihg}, we added the dark radiation self-interaction term that was derived analytically in~Ref.~\cite{Cyr-Racine:2015ihg}.
The parameter space is the standard $\Lambda$CDM,
plus five additional parameters $\lbrace{ \alpha, \xi, m_\mathrm{DM}, \alpha_\ell, \beta_\ell \rbrace}$ that encode the whole information about the interaction:
$\alpha$ is the amplitude of the scattering rate, $\xi=T_\mathrm{DR}/T_\gamma$ is related to the amount of dark radiation, $m_\mathrm{DM}$ is the dark matter mass, $\alpha_\ell$ and $\beta_\ell$ are coefficients depending on the angular dependence of the scattering rate, all defined in the Appendix.
We focus on a specific case of 3-particle interaction (i.e. fermionic dark matter + fermionic dark radiation + mediator),
which leads to a comoving scattering rate $\propto T^4$.
Note that there exist alternative dark sector set-ups not covered by the present analysis,
like e.g. the models discussed in Ref.~\cite{Buen-Abad:2015ova,Lesgourgues:2015wza} that lead to a scattering rate $\propto T^2$.
Finally, notice that the code treats dark radiation as a relativistic massless component.
Therefore, for the time being,
the analogy with eV sterile neutrinos has to be taken with some grain of salt.

{\it Impact on cosmological observables}---
Before investigating the cosmological bounds on the strength of the interaction,
let us review the effects of dark radiation dark matter interactions on the
cosmological observables at large scales.

\begin{figure}[h]
\includegraphics[width=\columnwidth]{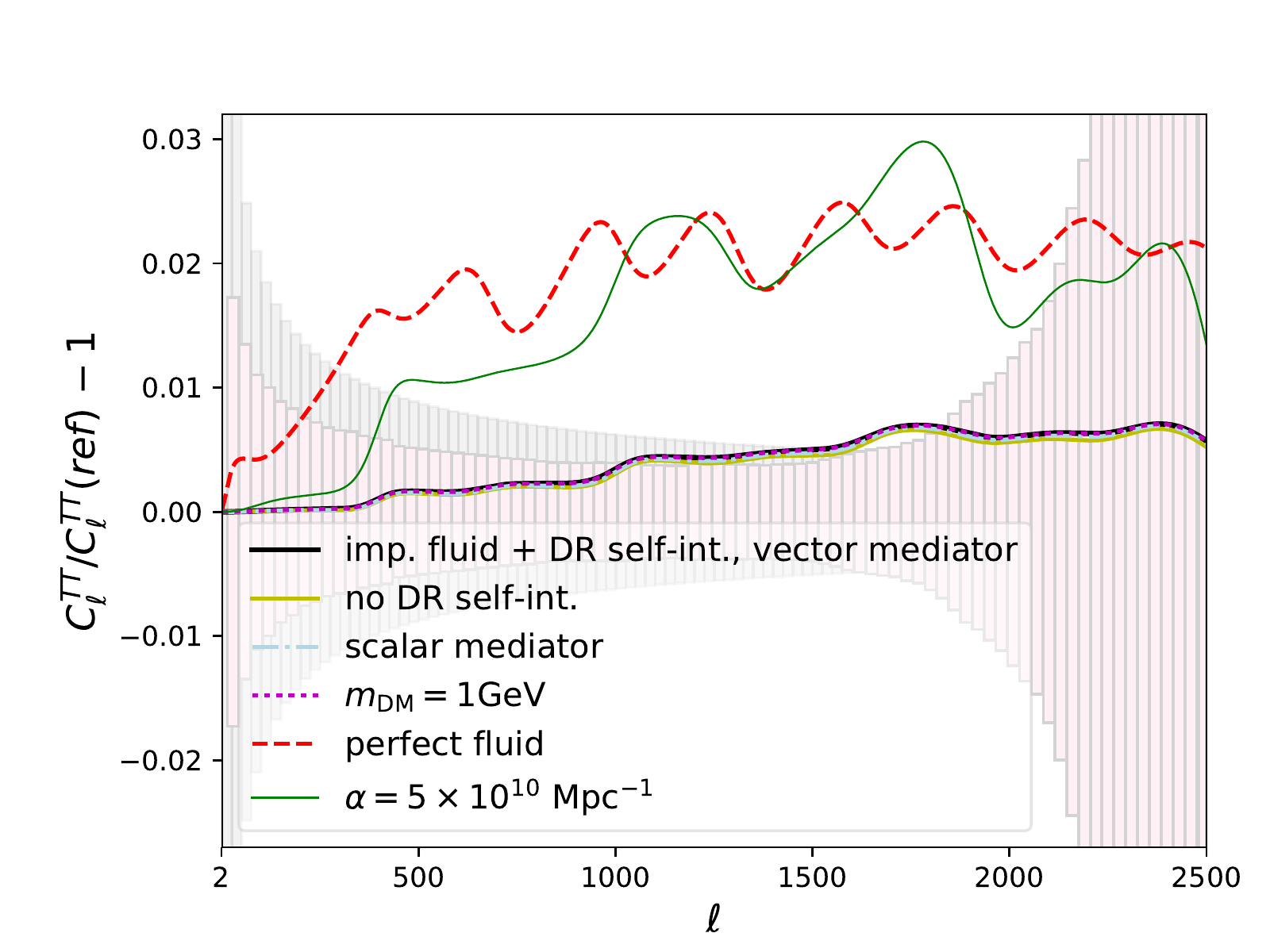}
 \caption{Residuals in the CMB temperature anisotropy power spectrum induced by changing the microphysics governing the interaction.
 The reference model is $\Lambda$CDM plus an equivalent number of extra neutrinos corresponding to $\xi=0.5$.
 The baseline interacting model (black thick solid line) is obtained with a dark matter mass $m_\mathrm{DM}=100$~GeV,
 dark radiation parameterized as an imperfect fluid, and a vector boson mediator.
 With respect to the baseline model, the yellow solid line does not include dark radiation self-interactions,
 the red dashed line represents a perfect dark radiation fluid,
 the cyan dot-dashed line replaces the vector mediator with a scalar mediator,
 the magenta dotted line has a reduced dark matter mass $m_\mathrm{DM}=1$~GeV,
 and finally the green thin solid line has a larger interaction constant $\alpha=5\times10^{10}$~Mpc$^{-1}$
 (in the other cases the constant is $\alpha=10^9$~Mpc$^{-1}$).
 The gray and pink shaded area depict, respectively, cosmic variance and Planck observational error
for a bin width $\Delta \ell=30$.
}
\label{fig:tt}
\end{figure}
Figure~\ref{fig:tt} shows the residuals of various interacting models based on different microphysics
with respect to a $\Lambda$CDM model with an equivalent number of extra neutrinos ($\Lambda$CDM+$\Delta N_\mathrm{eff}$, for the definition of $\Delta N_\mathrm{eff}$, Eq.~\ref{eq:neff} in the Appendix),
to get rid of background effects\footnote{
The main background effects due to an increased $N_\mathrm{eff}$~\cite{Archidiacono:2011gq,Archidiacono:2013fha,
Hou:2011ec,Bashinsky:2003tk} are
the delay of the radiation-matter equivalence,
the enhanced early Integrated Sachs Wolfe Effect,
the shift of the CMB acoustic peaks towards higher $\ell$
and the increase of the amount of Silk damping.}.
The spectrum of the baseline interacting model (black thick solid line)
is less suppressed wrt $\Lambda$CDM
than the $\Lambda$CDM+$\Delta N_\mathrm{eff}$ spectrum on small scales.
These scales enter the horizon while dark matter and dark radiation are still coupled,
i.e. the extra $\Delta N_\mathrm{eff}$ is not free-streaming.
In the absence of additional anisotropic stress, the gravitational source term of the photon oscillations
is not affected by the presence of extra radiation and the suppression is alleviated.
If the onset of dark radiation free-streaming is delayed after the time when all the scales contributing to CMB primary anisotropies have crossed the horizon,
then a second effect is also visible and it amounts to a phase shift
wrt $\Lambda$CDM+$\Delta N_\mathrm{eff}$ ($\alpha=5\times10^{10}$~Mpc$^{-1}$, green thin solid line).
Extra relativistic neutrinos induce a phase shift because they free-stream before photon decoupling.
Thus, neutrino perturbations are supersonic in the photon fluid
and can generate metric perturbations outside the sound horizon.
An additional dark radiation component with a late free-streaming would not induce the same phase-shift.
A third effect is induced by dark matter and dark radiation forming one single tightly coupled fluid,
so that dark matter is not pressureless anymore and its perturbations develop a fast mode~\cite{Voruz:2013vqa, Weinberg:2002kg}.
As a consequence, dark matter clusters less and, through gravitational interaction, 
this refelects into a suppression of the clustering of the baryon-photon fluid too.
The latter effect is clearly visible in the troughs at $\ell \sim 750,\,1400,\,2000$ for $\alpha=5\times10^{10}$~Mpc$^{-1}$ (green thin solid line): the troughs are located in correspondence of the odd peaks that reflect the compression phases and that are more suppressed in this scenario.
Finally, dark radiation can be described as a perfect fluid,
by truncating the Boltzmann hierarchy at $\ell=2$,
and setting the anisotropic stress $\sigma_\mathrm{DR}$ to zero (Eq.~\ref{eq:dr} in the Appendix).
In this case (red dashed line), the dark monopole and the dipole are prevented from transferring power to the higher order momenta.
Thus, neutrinos remain more clustered and enhance also the photon perturbations through gravitational coupling. This boosts the temperature spectrum on most causal scales.
Therefore, even if the interaction is confined to the dark radiation component
and does not affect Standard Model particles,
the free-streaming nature of the new relativistic component is relevant, and has to be carefully considered in model building.

\begin{figure}[h]
\includegraphics[width=\columnwidth]{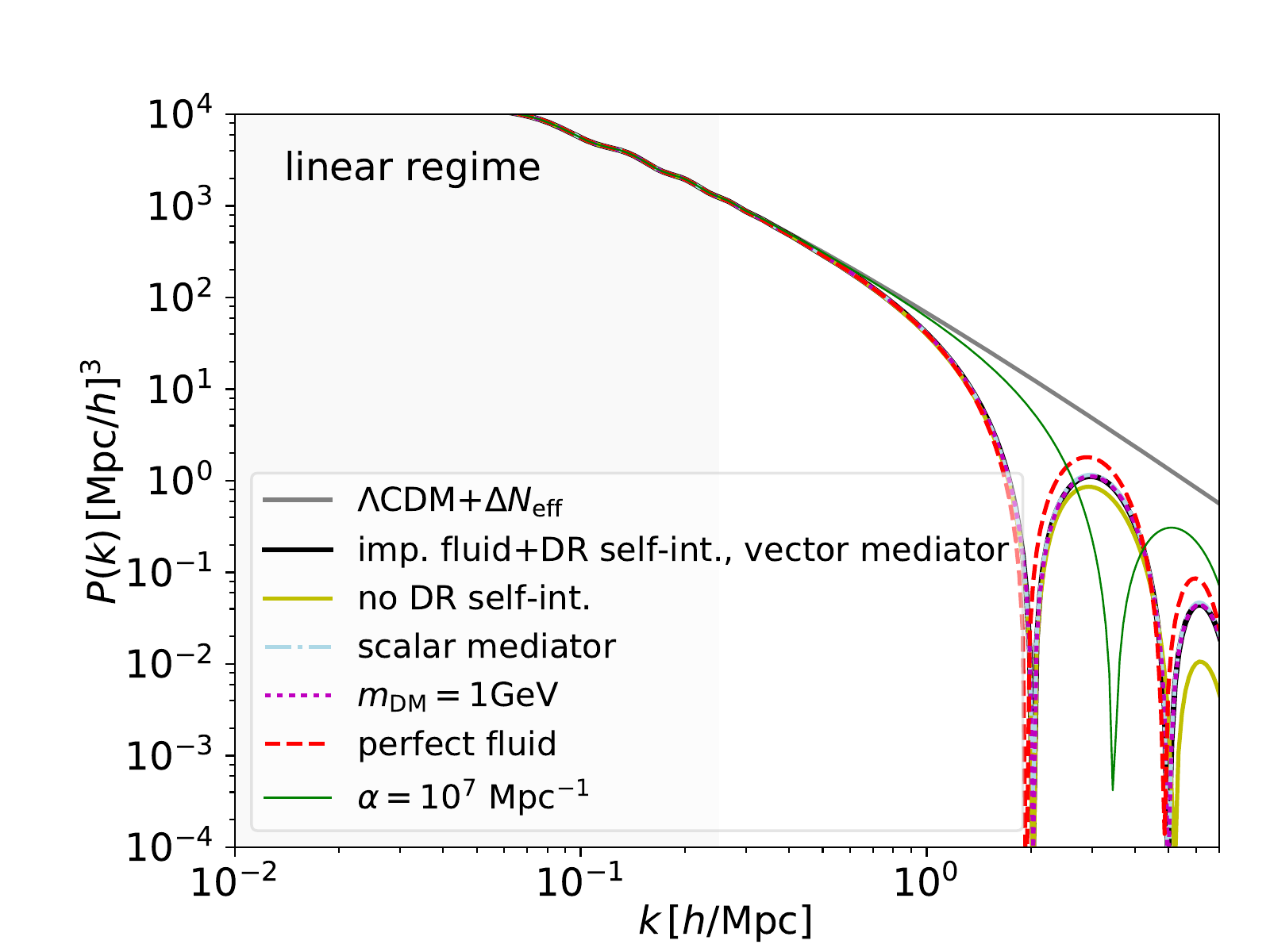}
\caption{Matter power spectrum for the same models discussed in figure~\ref{fig:tt},
but here $\alpha=10^8$~Mpc$^{-1}$ and we also show the baseline case for $\alpha=10^7$~Mpc$^{-1}$ (green thin solid line).
The gray shaded area defines the linear regime.}
\label{fig:pk}
\end{figure}
The matter power spectrum depicted in figure~\ref{fig:pk} reflects the fact that
dark matter perturbations cannot grow as long as dark matter is coupled to dark radiation.
As a consequence, the scales entering the horizon before kinetic decoupling are exponentially damped,
as in the case of warm dark matter~\cite{Lopez-Honorez:2017csg,Irsic:2017ixq}.
The difference compared to warm dark matter is that the interactions also leave an imprint
on the matter power spectrum through
acoustic oscillations of the dark plasma~\cite{Cyr-Racine:2013fsa}.
For smaller values of the coupling, the spectrum start deviating from $\Lambda$CDM+$\Delta N_\mathrm{eff}$ at smaller scales (green solid line).
However, the deviations are always located at non-linear scales.

Concerning the microphysics of the specific particle physics models behind our phenomenological parameterization,
we notice that the nature of the {\it mediator} (vector $\alpha_\ell=3/2$ or scalar $\alpha_\ell=3/4$, Eq.~\ref{eq:dr} in the Appendix)
does not have any significant impact (cyan dot-dashed line). 
Therefore, the constraints can be extended to particle physics models with different mediators~\cite{Binder:2016pnr}.
Moreover, removing {\it dark radiation self-interactions} (yellow solid line), which amounts in setting $\beta_\ell$ to zero (Eq.~\ref{eq:dr} in the Appendix), does not lead to an observable deviation both in the CMB temperature anisotropies
and in the matter power spectrum up to the first oscillation.
Thus, dark radiation self-interactions can be neglected.
Finally, decreasing $m_\mathrm{DM}$ from $100$~GeV to $1$~GeV does not have a significant impact (magenta dotted line). The reason is that in our parameterization the {\it dark matter mass} enters explicitly only in the expression for the dark matter sound speed $c^2_\mathrm{DM}$ (Eq.~\ref{eq:cs} in the Appendix), which is much smaller than the one of the tightly coupled DM-DR fluid, and has a minor impact on dark matter perturbations.

{\it Results}---
Given the considerations drawn in the previous paragraph, we have sampled only the $\Lambda$CDM+$\lbrace{ \alpha, \xi \rbrace}$ ($m_\mathrm{DM}=100$~GeV) parameter space, with a logarithmic prior on $\alpha$ in the range $[-3,14]$.
For too large values of $\alpha$, the cut--off in the power spectrum and the kinetic decoupling temperature are expected to be in conflict with Lyman--$\alpha$ data and galaxy formation. We do not study these bounds in the present paper and leave this study for future work. However this motivates the fact that we do not consider larger values of $\alpha$ in the present CMB analysis.

Our pipeline is based on {\sc MultiNest}~\cite{Feroz:2008xx,Feroz:2007kg,Feroz:2013hea}
in combination with {\sc MontePython}\footnote{\tt http://baudren.github.io/montepython.html}~\cite{Audren:2012wb}, and
interfaced with our modified version of {\sc class}. We use Planck 2015 polarization and temperature data (dubbed Planck low-$\ell$ P + high-$\ell$ TTTEEE \cite{Aghanim:2015xee}).
We have checked the conclusions that we have drawn above about the dark matter mass and about the nature of the mediator, do not affect the cosmological bounds on $\lbrace{ \alpha, \xi \rbrace}$.
We obtain $\xi<0.48$ and $\xi<0.44$ at 95\% c.l. for the imperfect fluid and the perfect fluid, respectively.
Translated into an equivalent neutrino number these values correspond to 
$\Delta N_\mathrm{eff}< 0.205$ (imperfect fluid) and 
$\Delta N_\mathrm{eff}< 0.144$ (perfect fluid) at 95\% c.l..
\begin{figure}[h]
\begin{tabular}{cc}
\includegraphics[width=0.5\columnwidth]{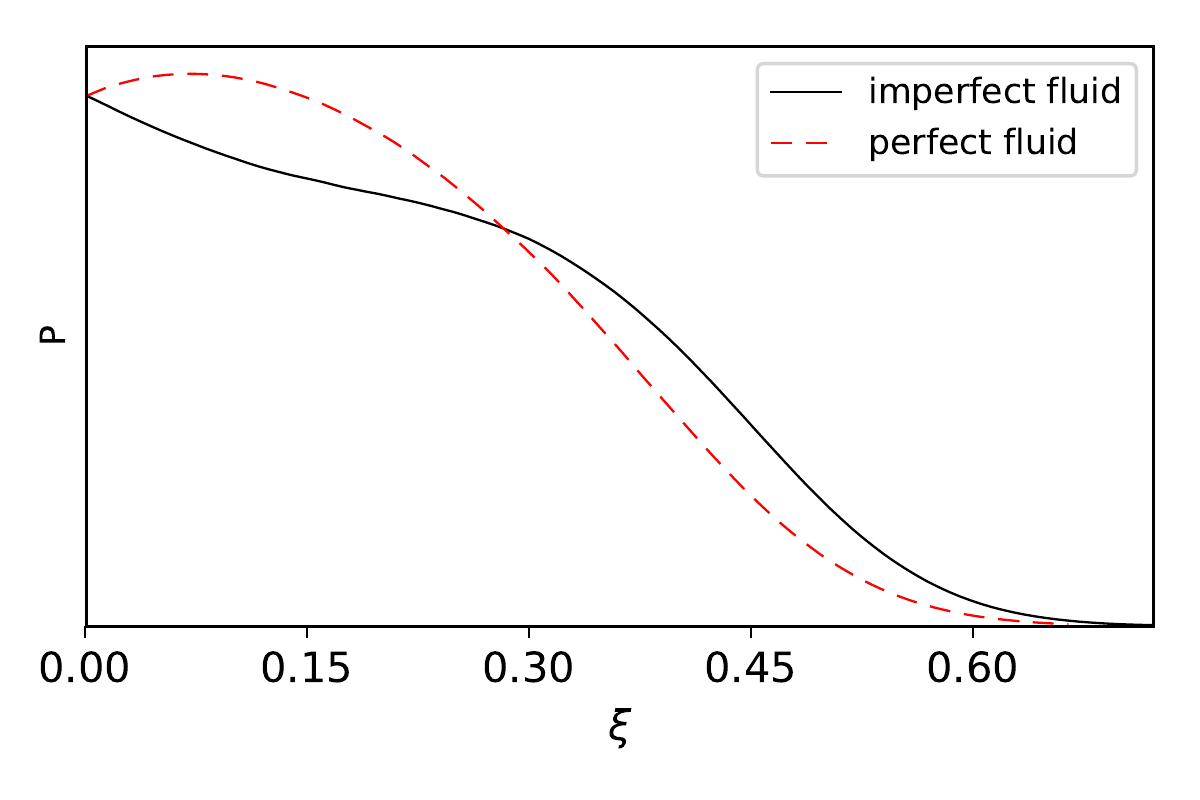}&\includegraphics[width=0.5\columnwidth]{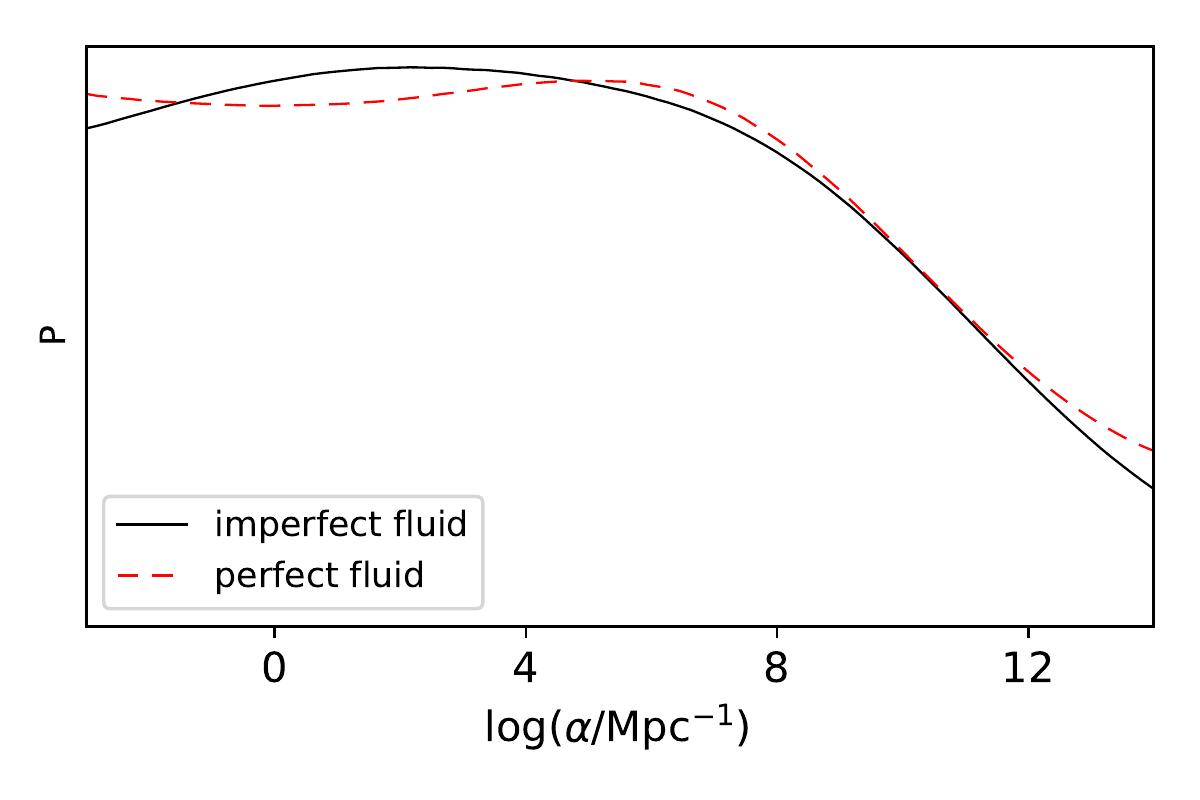}
\end{tabular}
\caption{
One dimensional marginalized posterior for $\xi$ (left panel) and $\alpha$ (right panel). 
}
\label{fig:1d}
\end{figure}

\begin{figure}[h]
\includegraphics[width=\columnwidth]{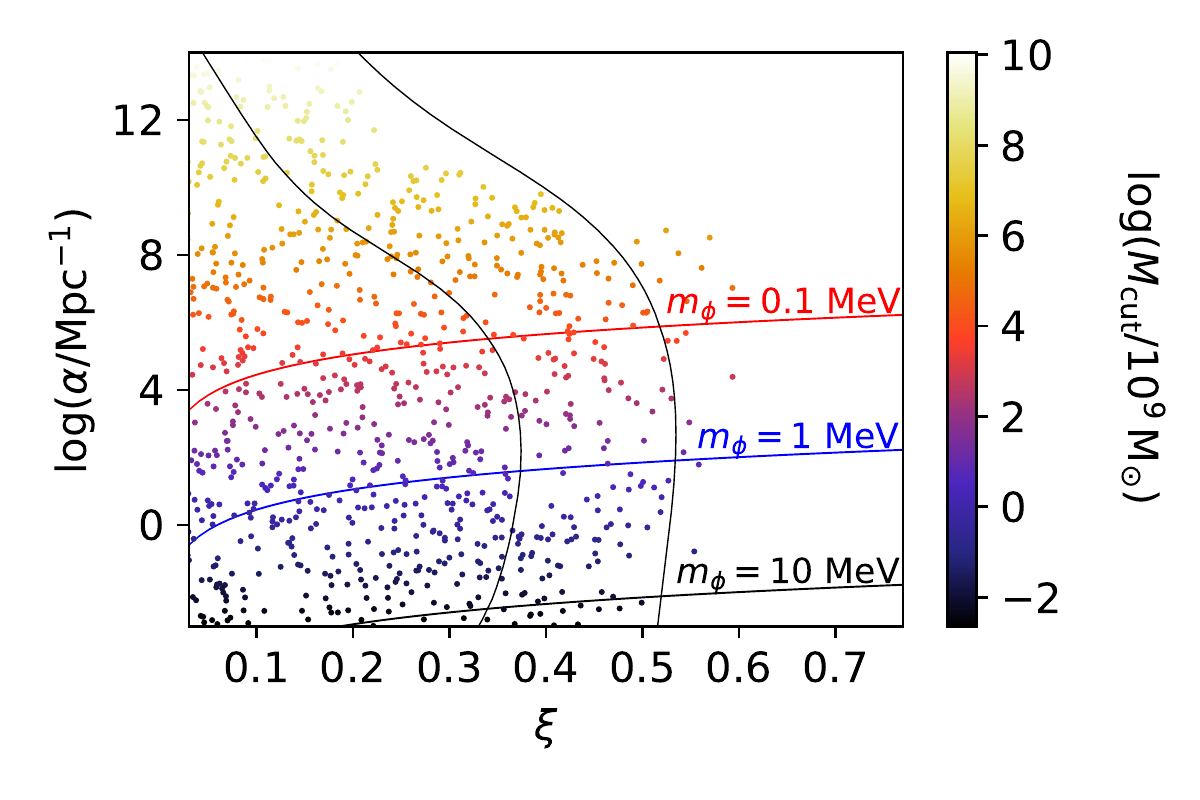}
\caption{
One and two $\sigma$ marginalized contours (black solid lines) in the plane $\left( \alpha, \xi \right)$ for the case of the imperfect fluid.
The scatter points are coloured according to the corresponding value of $M_\mathrm{cut}$.
The black, blue and red bands depict the regions of dark matter ($m_\mathrm{DM}=100$~GeV) thermal freeze-out for different values of the mass of the mediator ($m_\phi=0.1,1,10$~MeV, respectively), and assuming $g_\mathrm{DR}=g_\mathrm{DM}$.
}
\label{fig:3d_Mcut}
\end{figure}

{\it Connection with the small scale crisis}---
The constraints on $\alpha$ and $\xi$ can be converted into constraints on the
kinetic decoupling temperature $T_\mathrm{kd}$ of the dark sector:
\begin{gather}
T_\mathrm{kd}=\frac{0.062\,\mathrm{keV}}{10^{1/4}0.2} \left(\frac{\xi_\mathrm{kd}}{(4/11)^{1/3}} \right)^{1/2}
\left(\frac{61.7\,\mathrm{Mpc}^{-1}}{\alpha} \right)^{1/4}\nonumber\\
\left( \frac{\xi}{0.5} \right)^{1/2}.
\end{gather}
The kinetic decoupling temperature is then related to the minimal mass of sub-structures that can form in a galaxy $M_\mathrm{cut}$:
\begin{equation}
M_\mathrm{cut}=2.2 \times 10^8 \xi_\mathrm{kd}^3 \left( \frac{1\,\mathrm{keV}}{T_\mathrm{kd}}\right)^3 M_{\odot}.
\end{equation}
If $M_\mathrm{cut}$ is above $10^9\,M_{\odot}$ (i.e., $T_\mathrm{kd} \lesssim 0.6 \xi_\mathrm{kd}$~keV), the ``missing satellite'' problem is alleviated. 
Figure~\ref{fig:3d_Mcut} shows the one and two $\sigma$ marginalized constraints in the plane $\left( \alpha, \xi \right)$,
with the points coloured according to the values of $M_\mathrm{cut}$.
We can see that the parameter space allowed by CMB data is consistent with $M_\mathrm{cut} > 10^{9} M_{\odot}$\footnote{We
have assumed $\xi=\xi_\mathrm{kd}$, i.e. no entropy variation in the Standard Model sector and in the dark sector
between kinetic decoupling and today.
}.
Assuming dark matter thermal freeze-out and $g_\mathrm{DR}=g_\mathrm{DM}$,
we can derive the value of the mediator mass for each point in parameter space,
with the highest probabilty region corresponding to $0.1\,\mathrm{MeV} \leq m_\phi \leq 10\,\mathrm{MeV}$.
Finally, we find that the Bayes factor, i.e. the ratio of the bayesian evidence between the interaction model and the pure $\Lambda$CDM model,
is $1$, as expected from the behaviour of the one dimensional posterior of $\alpha$ (Figure~\ref{fig:1d}),
which is flat over many orders of magnitude ($\alpha = 0.001 - 10^8$).

{\it Discussion}---
We have studied the impact of a 3-particle interaction (fermionic dark matter + fermionic dark radiation + mediator)
on large scale cosmological observables,
namely the CMB temperature power spectrum and the matter power spectrum.
To this aim we have implemented a phenomenological description of the aforementioned scenario in {\sc class},
following the ETHOS parameterization of Ref.~\cite{Cyr-Racine:2015ihg}.

We find that variations of the theoretical model,
concerning, e.g., the dark matter mass, the vector or scalar nature of the mediator, and the presence of dark radiation self-interactions,
have little or no impact on the cosmological observables.
Whether dark radiation behaves as a perfect fluid with no anisotropic stress or resembles additional free-streaming neutrinos
does not affect the cut-off scale in the matter power spectrum.
However, for a given value of the amplitude of the scattering rate,
a dark fluid shows a deviation wrt $\Lambda$CDM+$\Delta N_\mathrm{eff}$
larger than a neutrino-like dark radiation.

We have performed a MultiNest Markov Chain Monte Carlo run, fitting the interaction model to the Planck CMB temperature and polarization data.
The region in parameter space corresponding to a cut-off mass in the range desirable for mitigating the ``missing satellite'' problem
is allowed by CMB data, with a bayesian evidence consistent with the one of a pure $\Lambda$CDM model.
In conclusion, this interaction model can induce a small scale cut-off in the halo mass function,
without diminishing the goodness of fit to large scale CMB data.

Finally, the parameter space allowed by CMB can be restricted by including information on clustering at non-linear scales.
In particular, Lyman-alpha data might pin down the temperature of kinetic decoupling.
However, notice that semi-analytical non-linear corrections (e.g. halofit) derived from N-body simulations of collisionless cold dark matter cannot be used in this context.

\bibliography{biblio}{}
\bibliographystyle{plain}

\section{Appendix}
We use the ETHOS parameterization proposed in Ref.~\cite{Cyr-Racine:2015ihg},
taking the dark radiation self-interactions into account.
The dark radiation Boltzmann hierarchy in Newtonian gauge looks as follows:
\begin{gather}
\dot{\delta}_\mathrm{DR}+\frac{4}{3}\theta_\mathrm{DR}-4\dot{\phi}=0,\\
\dot{\theta}_\mathrm{DR}+k^2\left(\sigma_\mathrm{DR}-\frac{1}{4}\delta_\mathrm{DR}\right)-k^2 \psi=\nonumber \\
\Gamma_\mathrm{DR-DM}\left(\theta_\mathrm{DR}-\theta_\mathrm{DM}\right),\\
\dot{\pi}_{\mathrm{DR},\ell}+\frac{k}{2\ell+1}\left((\ell+1)\pi_{\mathrm{DR},\ell+1}-\ell\pi_{\mathrm{DR},\ell-1}\right)=\nonumber \\
\left( \alpha_\ell \Gamma_\mathrm{DR-DM}+\beta_\ell \Gamma_\mathrm{DR-DR}\right) \pi_{\mathrm{DR},\ell}, \, \ell \geq 2.
\label{eq:dr}
\end{gather}
The dark matter perturbation equations are:
\begin{gather}
\dot{\delta}_\mathrm{DM}+\theta_\mathrm{DM}-3\dot{\phi}=0,\\
\dot{\theta}_\mathrm{DM}-k^2c_\mathrm{DM}^2\delta_\mathrm{DM}+aH\theta_\mathrm{DM}-k^2\psi=\nonumber \\
\Gamma_\mathrm{DM-DR} \left( \theta_\mathrm{DM} - \theta_\mathrm{DR}\right).
\end{gather}
In the equations above $\delta$ and $\theta$ are, respectively, the density and velocity dispersion perturbations,
$\pi_\mathrm{DR}=2\sigma_\mathrm{DR}$ with $\sigma_\mathrm{DR}$ the shear perturbation, $\phi$ and $\psi$ the gravitational potentials.
The dark sound speed $c^2_\mathrm{DM}$, defined as
\begin{equation}
c^2_\mathrm{DM}= \frac{T_\mathrm{DM}}{m_\mathrm{DM}} \left(1-\frac{\dot{T}_\mathrm{DM}}{3 a H T_\mathrm{DM}} \right)
\label{eq:cs}
\end{equation} 
depends on the heating rate and it amounts to a very small contribution for non-relativistic dark matter.
The terms on the right hand side of the dark matter and dark radiation dipole and of the dark radiation higher order momenta represent
the collisional integrals. 
The expression of $\Gamma_\mathrm{DR-DM}$ can be found in Ref.~\cite{Cyr-Racine:2015ihg}.
In the case of fermionic dark matter and dark radiation interacting through a massive mediator, the calculation of the matrix element of the scattering gives:
\begin{equation}
\Gamma_\mathrm{DR-DM}=- a \pi \frac{g^2_\mathrm{DR}g^2_\mathrm{DM}}{m_\phi^4} \left( \frac{310}{441} \right) n_\mathrm{DM} T^2_\mathrm{DR}, \label{eq:gdmdr}
\end{equation}
where $m_\phi$ is the mediator mass, $n_\mathrm{DM}$ is the dark matter density, $T_\mathrm{DR}$ is the dark radiation temperature,
parameterized as $T_\mathrm{DR} = \xi T_\gamma$, $g_\mathrm{DR}$ is the coupling of dark radiation to the mediator and $g_\mathrm{DM}$
is the coupling of the dark matter particle to the mediator.
In general, when the interaction rate is a power law of the temperature $\Gamma=a n_\mathrm{target} \langle \sigma v \rangle \propto T^n \propto (1+z)^n$, the inverse scattering rate is given by:
\begin{equation}
\Gamma_\mathrm{DM-DR}=\left(\frac{\rho_\mathrm{DR}}{\rho_\mathrm{DM}}\right) \left(\frac{2+n}{3}\right) \Gamma_\mathrm{DR-DM}~.
\end{equation}
In our case, equation (\ref{eq:gdmdr}) shows that this relation holds with $n=4$. We parametrise the scattering term as
\begin{equation}
\Gamma_\mathrm{DR-DM}=-\Omega_\mathrm{DM}h^2 \alpha \left(\frac{1+z}{1+z_d} \right)^4,
\end{equation}
where $z_d=10^7$ is a normalization factor related to the time of dark matter dark radiation kinetic decoupling
($z_d=10^7$ roughly corresponds to $T_\mathrm{kd} \sim 1$~keV).
In the large self-interaction limit, the DM and DR perturbation equations become stiff.
We don’t need to devise a tight-coupling approximation scheme,
since {\sc class} handles stiff systems without significant slow-down through its ndf15 implicit integrator.
However, we have checked that the spectra with and without the tight coupling approximation are consistent.

For the dark radiation self-scattering $\Gamma_\mathrm{DR-DR}$ we have assumed that $g_\mathrm{DR} = g_\mathrm{DM}$
as the simplest model would predict. In that case the self-scattering $\Gamma_\mathrm{DR-DR}$ is equal to $\Gamma_\mathrm{DM-DR}$.

Finally, $\alpha_\ell$ and $\beta_\ell$ that appear in the quadrupole and in the higher order momenta of the dark radiation hierarchy are the angular coefficients;
if the mediator is a vector (scalar) boson,  $\alpha_\ell=3/2$ ($\alpha_\ell=3/4$) for $\ell \geq 2$;
in the presence of dark radiation self-interactions, $\beta_\ell=1$.

The dark radiation component can be casted into an equivalent extra neutrino number as
\begin{equation}
\Delta N_\mathrm{eff}= \frac{g^*_\mathrm{DR}}{2}\frac{f}{7/8} \xi^4 \left( \frac{T_\gamma}{T_\nu} \right) ^{4},
\label{eq:neff}
\end{equation}
where $g^*_\mathrm{DR}$ is the dark radiation number of internal degrees of freedom (i.e. spin states, assumed throughout this work to be equal to two like for neutrinos),
$f$ is $7/8$ for fermionic dark radiation and
the factor $\frac{T_\gamma}{T_\nu}=\left( \frac{11}{4} \right) ^{1/3}$ has to be taken into account after neutrino decoupling.
If we assume that the dark sector decouples from the standard model at $T \sim 100$~GeV,
we obtain $\xi \sim 0.5$, which implies $N_\mathrm{eff} \sim 0.24$, still within Planck $1\,\sigma$ error.

\end{document}